\documentclass[12pt,preprint]{aastex} 
\usepackage{aastexug}
\shorttitle{Crab nebula spectroscopy}
\shortauthors{\v{C}ade\v{z} et al.}

\begin{document}

\title{Spectroscopy and 3D imaging of the Crab nebula}

\author{Andrej \v{C}ade\v{z}\altaffilmark{1}}
\affil{Department of Physics, Faculty of Mathematics and Physics, University of Ljubljana\\
Jadranska 19, 1000 Ljubljana, Slovenia}
\email{andrej.cadez@uni-lj.si}

\author{Alberto Carrami\~nana\altaffilmark{1}}
\affil{Instituto Nacional de Astrof\'{\i}sica, \'Optica y Electr\'onica, Luis Enrique Erro 1,\\ Tonantzintla, Puebla 72840, M\'exico}
\email{alberto@inaoep.mx}

\author{Simon Vidrih\altaffilmark{1}}
\affil{Department of Physics, Faculty of Mathematics and Physics, University of Ljubljana\\
Jadranska 19, 1000 Ljubljana, Slovenia}
\email{simon.vidrih@fmf.uni-lj.si}

\altaffiltext{1}{Visiting Astronomer, Observatorio Astrof\'{\i}sico Guillermo Haro, Cananea, Sonora.}

\begin{abstract}
Spectroscopy of the Crab nebula along different slit directions reveals the 3 dimensional structure of the optical nebula. On the basis of the linear radial expansion result first discovered by Trimble~\cite{c1}, we make a 3D model of the optical emission. Results from a limited number of slit directions suggest that optical lines originate from a complicated array of wisps that are located in a rather thin shell, pierced by a jet. The jet is certainly not prominent in optical emission lines, but the direction of the piercing is consistent with the direction of the X-ray and radio jet. The shell's effective radius is $\approx 79$ seconds of arc, its thickness about a third of the radius and it is moving out with an average velocity $1160\,\mathrm{km\,s^{-1}}$.
\end{abstract}


\section{INTRODUCTION}

The Crab nebula was an object of  intense interest even before its pulsar was discovered in radio by Staelin and Reifenstein~\cite{c14} and in the optical by Cocke, Disney, and Taylor \cite{c2}. Trimble \cite{c1} first measured proper motions of nebular components and, combining them with spectroscopic data taken by Guido M\"unch~\cite{c15}, discovered that they all point to a very small, almost point-like origin in the past.  Velusamy, Roshi, \& Venugopal \cite{c3} obtained a similar result from radio images of the nebula. Both observations are consistent with an outward pointing velocity field with the proper motion magnitude proportional to the distance from the origin. Trimble \cite{c1}, Wyckoff and Murray \cite{c4}, Caraveo and Mignani \cite{c5} pointed out that the position of this origin is displaced with respect to that of the pulsar. Furthermore the proper motions of nebular components are about 10\% faster than expected if the nebular debris were expanding freely. In radio images the linear expansion is not as obvious and shows some anomalies further away from the pulsar. More recently, attention has focused to phenomena connected to Crab's high-energy emission, its jet and to the very rapid expansion of inner wisps discovered optically by Hester et al. \cite{c6} and in radio by Bietenholz et al. \cite{c7}. It has become increasingly clear that the Crab pulsar and its nebula are connected through a multi layer structure of different predominant energy ranges. To understand the mutual interrelationship of these layers, it is useful to have a three dimensional picture of at least part of the structure.

The first attempts to understand the 3D structure of the Crab nebula go back to Chevalier and Gull \cite{c13} who studied the nebula in the light of all 13 important visible emission lines by using narrow band filter imaging. They found a structure "consisting of dense filaments from which thin clumpy sheets of gas fan out radially". Clark et al. \cite{c19} studied spectroscopically the radial velocity field across the nebula to investigate the nebula 3D properties and find the line emitting region to be a thick hollow shell composed of bright inner and faint outer components. They also see the filaments as generally circumferential, but radial "spokes", connecting the inner and outer shells. They conclude that the nebular synchrotron emission is confined within the inner brighter shell. A comprehensive attempt to understand the geometry, composition and mass of the Crab nebula has been done by MacAlpine et al. \cite{c20} and \cite{c21}. They combine interference filter imaging and long-slit spectroscopy to obtain information on composition and distribution of gas in the nebula. On the basis of a few selected spectra and using model calculations MacAlpine et al. \cite{c20} deduce that collisional contributions to observed exceptionally strong helium lines are insignificant. Using further model assumptions they estimate the combined mass in the filaments to be in the range 6-9 $M_\odot$. On the basis of interference filter and Fabry-P\' erot imaging Lawrence et al. \cite{c11}
interpret the [OIII]$\lambda 5007$ emission to have a spatial relationship to a "high-helium torus".

The possibility to make a 3D representation of the line emitting regions in the Crab nebula is offered by the proper motion result of Trimble and others. On the basis of this result one is led to speculate that all the three components of the velocity vector field describing the motion of nebular components point in the outward radial direction since the motion is ballistic. According to this assumption the speed of this motion is proportional to the distance from the origin, i.e. we assume:
\begin{equation} \label{eq:trimble}
\mathbf{v} = \mathbf{r}/T \quad ,
\end{equation}
where $\mathbf{v}$ is the velocity vector at the position $\mathbf{r}$ measured from the center of the nebula, and $T$ is time elapsed. If the ballistic assumption was fully satisfied, $T$ would be the time since the explosion in 1054, but, as Trimble's proper motion results show the possibility of some acceleration of the debris in the past, we consider $T$ as a model parameter that may differ somewhat from the time since the explosion. Observing spectroscopically a small patch of the nebula at the angular position $(\xi,\,\eta)$ from the pulsar, one expects to find multiple Doppler shifted emission lines, each Doppler shift corresponding to a different radial velocity component ($v_z$) of the cloud. According to (\ref{eq:trimble}), such a component corresponds to a line of sight coordinate\- $z = ({v_z} - {v_{zc}}) T$, while the other two components of the vector $\mathbf{r}$ are $x = (\xi - \xi_c) D$ and $y = (\eta - \eta_c) D$, where $ v_{zc}$ and $(\eta_c, \xi_c)$  are the velocity and (angular) coordinates of the center of the cloud, while $D$ is the distance to the nebula. Locating line emitting regions in this way, one can construct a tentative 3D image of (line emitting) nebular components. The only parameter critical for the shape of such a reconstructed image is the ratio $T/D$. We choose to consider it as a free parameter and construct 3D models of the nebula for different values of it. We find it remarkable that it is possible to find a value for this parameter such that the 3D model of the nebula in  the radiation from the strongest optical emission lines ([NII], [SII], and  H$_\alpha $) appears as embedded inside a relatively thin, pierced spherical shell with a well defined inner radius. We consider this particular ratio $T/D$ of some interest, yet, we would like to emphasize that its implications for the distance to the nebula depend strongly on the ballistic assumption.

\section{DATA}

The spectra of the Crab nebula were obtained on the 18$^{th}$ of January 2001 (slit orientation $0^\circ$ and $90^\circ$, high resolution), on the 4$^{th}$ ($0^\circ$ and $90^\circ$, low resolution ), 5$^{th}$ ($18^\circ$, $36^\circ$, $54^\circ$ and $72^\circ$, low resolution), 6$^{th}$ ($108^\circ$, $126^\circ$, $144^\circ$ and $162^\circ$, low resolution), 7$^{th}$ ($36^\circ$, $72^\circ$, $108^\circ$ and $144^\circ$, high resolution) and 8$^{th}$ ($18^\circ$, $54^\circ$, $126^\circ$ and $162^\circ$, high resolution) of January 2002 with the 2.12 m telescope of the Observatorio Astrof\'{\i}sico Guillermo Haro in Cananea, Mexico, equipped with the Boller and Chivens spectrograph. All observations were done under photometric conditions. Spectra were taken in a low resolution mode of 1.6\AA~per pixel and a 250$\mu$m ($\approx$ 4 pixels $\approx$ 2 arc seconds) wide slit, and in a high resolution mode of 0.43\AA~per pixel and a 50$\mu$m slit. The spatial FWHM of the pulsar profile was varying between 2 and 3 arc seconds during observations and was thus wider than the projected width of the slit. The slit length was 560 pixels$\times$0.46 arc seconds per pixel in both cases, however, all the pixels were not exposed properly (see Fig.~\ref{Oba036x}). In the first case the effective Doppler width of the slit was $300\,\mathrm{km\,s^{-1}}$ and in the second it was about $16\,\mathrm{km\,s^{-1}}$. Five spectra (20 min exposure each) were taken with the low resolution and three (30 min exposure each) with the high resolution for each of the 10 slit orientations ($\psi_s~=~0^\circ,~18^\circ,~36^\circ\dots 162^\circ$ with respect to the EW orientation). The low resolution spectra cover the spectral region between 5400\AA \ and 7000\AA \, and give an average signal to noise ratio (S/N) of more than 200 per pixel. The high resolution mode covers the region between 6400\AA \ and 6840\AA \, and yields an average S/N of only about 2. Thus, in the low resolution mode the nebular continuum is clearly detected with the S/N of up to about 40 per pixel, while in the high resolution it is just below the detection threshold. As an example, reduced (bias subtraction, sky flat, median average) high and low resolution spectra for the slit orientation $36^\circ$ are presented in Fig.~\ref{Oba036x}. In Fig.~\ref{Super036} we also show a rescaled superposition of the two, where the high resolution spectrum is colored in red and the low resolution one in green. A trace of the spectrum  (at about 75 arc sec from the pulsar, below in Fig.~\ref{Oba036x}b) is shown in Fig.~\ref{Identi}. The rest frame wavelengths of a number of known nebular spectral lines are marked in gray and the Doppler shifted  lines of nitrogen, sulfur, hydrogen, oxygen, and helium are marked in blue ($-1047\,\mathrm{km\,s^{-1}}$) and red ($+1729\,\mathrm{km\,s^{-1}}$). The scale on the ordinate axis is in photon counts per pixel per 20 minutes. Finally, Fig.~\ref{Continuum0036}. shows the continuum trace along the $\psi_s~=~36^\circ$ slit  at approximately 6000\AA , i.e. in the flat region between the HeI and [OI] lines. The dot shows the position of the pulsar, where the continuum value interpolated along the dotted line is calibrated with respect to previous measurement \cite{c8}. Some features in this and other continuum traces can be recognized as belonging to stars that can be identified in the image of the nebula (the peak just next to the right of the pulsar belongs to the star $\approx$ 6 arc seconds away). We use this information to secure  accurate positions of the slits with respect to the nebula. The positions  thus derived are shown in Fig.~\ref{Slits}. The (small) corrections with respect to desired positions, obtained by these data have been taken into account in further data analysis of low resolution spectra. This procedure could not be applied for the high resolution spectra because of the low S/N. In this case we assume that the slits crossed the pulsar accurately and that the transparency of the sky was constant. We expect that the respective error in position and in flux is comparable to the shifts and calibration factors obtained in low resolution spectra ($\pm 4$ arc seconds in position and about $\pm10$\% in flux). In constructing images we use the more precise position data from low resolution spectra and use high resolution spectra to reduce the velocity calibration errors of the low resolution spectra.

\placefigure{Oba036x}
\placefigure{Super036.ps}
\placefigure{Identi}
\placefigure{Continuum0036}
\placefigure{Slits}

An analysis of all the spectra confirms the impression apparent from Fig.~\ref{Identi}, that the intense [NII] and  [SII] doublets, H$_\alpha$ line and also the weak [OI] doublet always occur at a common red-shift, relative intensities varying across the nebula. The forbidden [NII]5755 line is almost always clearly absent, but in some instances its presence may be masked by the double atmospheric feature just to the left and to the right of the red [NII]5755 mark (compare also Fig.~1). Only a close scrutiny revealed the presence of this weak line in the $36^{\circ}$ spectrum at the red-shift of the strongest [NII] line. It is quite difficult to automatically obtain good estimates for the ratio $(I[6548]+I[6583]) / I[5755]$ from these spectra, since the continuum is not well defined in this region because of the mentioned presence of weak atmospheric lines. Checking spectral traces by hand and judging the intensity of the $I[5755]$ line by the value at the pixel corresponding to the velocity of the $I[6548]$ and $I[6583]$ line, we estimate some values for this ratio between $185\pm 35$ ($I[6548]+I[6583]=259$photon counts/pixel), $210\pm 35$ ($I[6548]+I[6583]=487$photon counts/pixel) and $310\pm 100$ ($I[6548]+I[6583]=193$photon counts/pixel) for traces, where the $I[5755]$ line is visible. This value is consistent, even somewhat lower than that found by Davidson et al. \cite{c18}. 

Among the nebular lines observed in our spectra, we take only the five most intense as characterizing optical emitting regions: [NII] ($i~=~1,~2$), H$_\alpha$ ($i~=~3$) and [SII] ($i~=~4,~5$). To determine their common Doppler shift and strength we proceed as follows: for each slit orientation we scan the spectra along all slit length positions $m$ ($10\,\textrm{pixels}<m<537\,\textrm{pixels}$ for high resolution spectra and $43\,\textrm{pixels}<m<545\,\textrm{pixels}$ for low resolution spectra; the slight misalignment of the slit with respect to the CCD axis is taken into account) and obtain raw spectral traces $S_r(\lambda , m \vert \psi_s)$. These are flux calibrated by making use of the fact that all slit positions cross the pulsar. Thus we set the 6000$\,\mathrm{\AA}$ interpolated nebular continuum at the position of the pulsar to the same reference value for all spectra and analyze all spectra in units of this reference value (see Fig.~\ref{Continuum0036})\footnote{We estimate that the relative error of this recalibration is of the order of $\pm3\%$, about the size of the dot in Fig.~\ref{Continuum0036}.}. Pure line spectra $S_l(\lambda, m \vert \psi_s)$ are then obtained by subtracting the continuum. It is automatically defined for each trace ($m$) as the  best $n^{th}$ (n=10) degree polynomial fitting the points $(\lambda_i^{(c)}, S_r(\lambda_i^{(c)}, m \vert \psi_s))$, where $\lambda_i^{(c)}$ are determined by inspection of the whole spectrum at the wavelengths free of spectral lines\footnote{The recalibration of the spectrum and continuum subtraction is possible only for the low resolution spectra where the continuum is observed with sufficient signal to noise ratio. In high resolution no continuum is observed and no subtraction is applied, therefore, no common calibration intensity is present.}. Lines and their common Doppler shifts are finally identified defining a Doppler velocity dependent weight for each of the five lines in the spectrum as:
\begin{equation}
 w_i(v,m)~=~\int S_l(\lambda ,m)~e^{-{{\lambda - \lambda_i(1-v/c)}\over {2
 \sigma^2}}}d\lambda \quad .
\end{equation}
Choosing the appropriate value for the parameter $\sigma$ we form the velocity weight function:
\begin{equation}
\Omega(v)~=~w_1(v) w_2(v) ~+~ w_c w_3(v) ~+~ w_4(v) w_5(v)
\end{equation} 
and find on the interval $-2000\,\mathrm{km\,s^{-1}}<v<2000\,\mathrm{km\,s^{-1}}$ the velocity $v_{max}$  at which $\Omega(v)$ is maximal. Such a velocity weight function is large only if both lines of doublets are present and it has been found experimentally on our data sample (some 55,000 spectral traces) that its maximum gives a velocity that can be considered the common velocity of the three components (H, N and S). The characteristic strength $w_c$ introduced with the H$_\alpha$ line is chosen so that the contribution of this line to the velocity weight function is in the average comparable to the contribution of the two doublets. The Gaussian lines, Doppler shifted to $v_{max}$ are then subtracted from $S_l(\lambda,m)$ and the procedure is repeated on the remaining spectrum until $\Omega(v )$ no longer shows significant maxima. The important parameter $\sigma$, chosen as constant in the reduction of a given spectrum, is selected before the automatic procedure starts. This is done through trial and error until most randomly chosen line multiplets subtract best in a single subtraction. In the low resolution case the best $\sigma$ value is found to be essentially the effective slit width. This is expected, since no lines show evidence of being resolved. In this case the subtractions of all the five lines belonging to a common velocity multiplet usually leave only little more than the expected photon counting statistical noise, except in few cases where different  Doppler shifted components of H$_\alpha$ and [NII] overlap. The weights $w_i(v_{max},m)$ form what we call reduced spectra and the set of the five weights for a given $v_{max}$ and $m$ is called here a velocity component. Three examples of successive subtractions are shown in Fig.~\ref{OdstGrobi}. In this way we detected 43843 velocity components in all the 10 low resolution spectra, in the average 8.5 components per spectral trace. A great majority of these are of no consequence, since their intensity is negligible. $80$\% of light is contributed by only 5300 (12\%) of different velocity components, and 20\% account for 90\% of light. In other words, 8770 velocity components (1.7 per spectral trace) account for 90\% of light. The remaining 80\% of detecte velocity components contribute little, so they are barely seen in the reconstruction. Some of them belong to line wings that are not subtracted during the first run, to weak independent lines, to residues of imperfect subtraction which are due to either photon counting noise or small errors in wavelength calibration, but also due to the fact that different lines sometimes overlap and their components become confused. It is difficult to precisely judge the relative contribution of these effects. We believe they are best illustrated in Fig.~\ref{OdstGrobi}, which shows typical examples for reasons of imperfect subtraction to the level of a few percent and in Fig.~\ref{PraviInSinteticni}, where the observed spectrum and its reconstruction from detected velocity components can be compared.

\placefigure{OdstGrobi}
\placefigure{PraviInSinteticni}

The high resolution spectra clearly show lines of different resolved widths. A drastic example of both an unresolved [SII] doublet at $-115\,\mathrm{km\,s^{-1}}$ and a $150\,\mathrm{km\,s^{-1}}$ broad [SII] doublet at $814\,\mathrm{km\,s^{-1}}$ is found in the spectral trace along the line in Fig.~\ref{Oba036x}a and is shown in Fig.~\ref{OzkiInSiroki}\footnote{Note that the Doppler width of thermal broadening for hydrogen at the temperature 8000K is $11.5\,\mathrm{km\,s^{-1}}$, so our spectral resolution is not sufficient to show line broadening due to this effect. We expect that at this level of resolution the width of the line tells mainly about the spatial extent of the radiating gas.}. In this case it would be advantageous to take $\sigma$ as a free parameter for each system of lines. However, since most lines are neither as narrow nor as wide as the example shown here and the S/N is quite low, we usually choose a moderate $\sigma = 40\,\mathrm{km\,s^{-1}}$  which subtracts most lines rather well. Only the broadest and strongest lines produce additional sidebands next to the main peak in reconstruction. A clear example of such a broad line is clearly seen in Fig.~\ref{FiniPr036} on top of the $\psi=162^\circ$ and also on top of the $\psi=0^\circ$ spectrum.

\placefigure{OzkiInSiroki}

Spectra reconstructed from velocity components are shown in Figs. \ref{Grobi036} and \ref{FiniPr036}, where the intensity of the weights is color coded as indicated by color triangles in the middle. The intensity is defined as $I$=${1\over I_{max}(\psi)}(w_1+w_2+w_3+w_4+w_5$), where $I_{max}(\psi)$ is the maximum intensity for the given reduced spectrum. In units of interpolated continuum intensity at the pulsar position $I_{max}(\psi)$ are: 42.5, 22.4, 61.7, 41.7, 38.3, 46.0, 30.1, 43.0, 57.5, 74.0 in order from $\psi=0^\circ$ to $\psi=162^\circ$. The colors are assigned with respect to the "average bright spot mixture of intensities", which is coded gray (center of the triangle, $\lbrace \mathrm{H:N:S} \rbrace$=$\lbrace 0.18: 0.42: 0.40\rbrace$), while the edges of the triangles correspond to pure H, N or S as indicated on top in Fig.~\ref{Grobi036}. Comparing Figs. \ref{Grobi036} and \ref{FiniPr036} (compare also Fig.~\ref{Oba036x}) one notes the advantages and disadvantages of high and low resolution. The dominating features are clear and accurately correspond in both figures. However, the global features of the radiating gas distribution are better appreciated in low resolution due to higher signal to noise ratio. On the other hand, only the high resolution shows that  the radiating gas is organized in exquisitely fine sheets, tubes or blobs.
\placefigure{Grobi036}

Reduced spectral intensities are considered the line intensities at the position:
\begin{equation}
\xi=\widetilde m \cos \psi+\Delta \xi(\psi),~~~~~\break \eta=\widetilde m \sin
\psi+\Delta \eta(\psi),~~~~~\break
v=v_{max}+\Delta v(\psi) 
\label{Coord}
\end{equation}
of the 3D image and are stored accordingly. Here $\widetilde m = \bigl [m-m_0(\psi)\bigr ]\times 0.46$ arc sec/pixel and $m_0(\psi)$ is the value of $m$ in the spectrum $S_r(\lambda,m)$ where the spectral trace belonging to pulsar light is strongest. $\Delta \xi(\psi)$ and $\Delta \eta(\psi)$ are coordinates of the vector connecting the slit position ($m_0(\psi)$) with the center of the pulsar in the image; see the lower part of Fig.~\ref{Slits} (for $\Delta \xi(0)$ and $\Delta \eta(0)$). The $\Delta \xi$, $\Delta \eta$ values are of the order of 2 arc seconds, except for the slits $\psi=72^\circ$ and $162^\circ$, where the star next to the pulsar was mistaken for the pulsar. The velocity calibration error of low resolution spectra $\Delta v(\psi)$ was determined by graphically correlating low resolution and high resolution synthetic spectra, assuming that high resolution spectra velocity error is negligible. The values found are between $-180\,\mathrm{km\,{s^{-1}}}$ and $0\,\mathrm{km\,{s^{-1}}}$. From high resolution spectra $\Delta \xi (\psi)$ and $\Delta \eta (\psi)$ and $\Delta v(\psi)$ can not be obtained, but the comparison with low resolution spectra suggests that the pointing error was not larger than the one reached in taking low resolution data. 
\placefigure{FiniPr036}

\section{\label{image}IMAGES}

On the basis of the spectral information alone we first reconstruct a) the  continuum image of the nebula at 6000 \AA \ and b) the image of the nebula in the light of the five strong lines. They are shown together with the optical, radio and X-ray image in Fig.~\ref{FourImages}. The black and white image corresponding to the 6000$\,\mathrm{\AA}$ continuum (bottom left) is formed by assigning the values for the continuum intensity at measured points $(\xi_k,~\eta_k)$ and values in between are assigned by linear interpolation, i.e. if $S_r(\lambda,\Delta m+m_0(\psi_s)\vert \psi_s)=S_1$ and $S_r(\lambda,\Delta m+m_0(\psi_s)\vert \psi_s+\Delta \psi)=S_2$, then $S_r(\lambda,\Delta m+m_0(\psi_s)\vert \psi_s+\delta \psi)=S_1+(S_2-S_1){\delta \psi \over \Delta \psi}$, where $\Delta \psi~=~18^\circ$ and $0\leq \delta \psi < \Delta \psi$, and the coordinates $\xi$ and $\eta$ belonging to points $m~=~\Delta m+m_0(\psi)$, $\psi=\psi_s+\delta \psi$ are calculated according to eq.~\ref{Coord}.

The emission lines image (bottom right) is formed in a similar fashion, but in color so that the green intensity is proportional to the intensity of nitrogen regardless of velocity, the red component comes from H$_\alpha$ and the blue from sulfur. The color contributions are normalized so that the strongest line of any color gives the maximum display value for this color (255). 

In order to appreciate the level of detail that such synthetic images can give, we also display in Fig.~\ref{FourImages}  a blend of the visual image and the X-ray image (top left), the radio image (top center) and an enhanced visual image (top right) of the nebula. We find the continuum image reminiscent of the X-ray image, while the N, H, S image clearly highlights the filament structure embedded in the nebula, as does the radio image.

\placefigure{FourImages}

The 3D image can be understood on a computer screen if the structure is rotated and displayed as a movie (which can be obtained in the electronic version of this paper; see also Fig.~\ref{Movie}) which is based on low resolution data. The color code is the inverse\footnote{cyan$\to$red, yellow$\to$blue, magenta$\to$green} of that used in Fig.~\ref{Grobi036} and the intensity is coded as density of radiating fog multiplied by $\sin (\theta)$, where $\theta$ is the angle between the $Z$ axis and the direction from the coordinate system origin to the observed cloud. This is to compensate for the difference in sampling density. The axes $X$ and $Y$ are oriented as in Fig.~\ref{Slits} and their length is $90~\mathrm{arc~sec}$, while the $Z$ axis points away from the observer and its length is $1300~\mathrm{km/s}$. When observing this animation, it becomes quite apparent that the structure is shell like and that it is possible to adjust the $T/D$ value so that it looks round when rolled on the screen. Therefore, we fit a spherical shell to the distribution of line radiators by minimizing the action:
\begin{equation}
A[\xi_c,\eta_c,v_c,\Delta,\alpha]= {1\over \sum_k u_k}
\sum_{k} \left(\sqrt{(\xi_k-\xi_c)^2+(\eta_k-\eta_c)^2+\alpha^2
(v_k-v_c)^2}-\Delta\right)^2 u_k \quad .
\end{equation}
\placefigure{Movie}
Here $u_k$ is the weight of the $k$-th radiating blob formed from line intensities\footnote{For high resolution spectra  $u_k=\sum_{i=1}^5 w_i(\xi_k,\eta_k,v_k)$. In the case of low resolution spectra the inspection of Figs.~\ref{Grobi036} and~\ref{FourImages} makes obvious that the $\rho - v$ plane of the nebula has not been scanned exhaustively, some waining filaments having clearly escaped detection since the slit of the spectroscope is not long enough to reach to the farthest radiating regions. We note, however (Fig.~\ref{Grobi036}), that the intensity generally decreases with the distance from the shell, so that weighting brighter clouds more heavily than the dimmer ones might not displace the center of the nebula, but it does overcome the geometric selection problem. Therefore, we either exclude all clouds with less than a few percent of maximum intensity or by weighting $u_k$ proportional to $w_i^2$. Both methods give similar results which differ only by about 1\% in $\Delta$,  by about 5 arc seconds in $(\xi_c,\eta_c,\alpha v_c)$ but almost 30\% in $\alpha$ and agree with those from the high resolution spectra.} $w_i(\xi_k,\eta_k,v_k)$ and the variational parameters $(\xi_c,\eta_c)$, $v_c$, $\Delta$, $\alpha$ are the position of the center of the shell with respect to the pulsar, the velocity of the shell center with respect to the observer, the effective shell radius and $\alpha=T/D$ respectively. In the case of the low resolution the best fit was obtained for the following parameter values: $(\xi_c,\eta_c,\Delta, v_c)=(10.4\,\mathrm{arc\,sec}, 17.8\,\mathrm{arc\,sec}, 79.1\,\mathrm{arc\,sec}, 198\,\mathrm{km\,s^{-1}})$ with the action value of (13.2 $\mathrm{arc\,sec}$)$^2$. In the high resolution case the parameter values are: $(\xi_c,\eta_c,\Delta, v_c)=(11.3\,\mathrm{arc\,sec}, 13.8\,\mathrm{arc\,sec}, 74.5\,\mathrm{arc\,sec}, 189\,\mathrm{km\,s^{-1}})$ for the action value of (10.1 $\mathrm{arc\,sec}$)$^2$. The best fit parameters were used to plot projections of the shell on the $v-\rho$  planes in Figs.~\ref{Grobi036} and~\ref{FiniPr036}. The dynamical center of the nebula as found by our fit $(\xi_c,\eta_c,\alpha, v_c)$ is plotted as a cross in Fig.~\ref{FourImages}; both the fine resolution and the low resolution spectra obtain center within the box defined by the size of the cross (about 10 arc seconds).
\placefigure{144Sat}
The best value for $\alpha$ is $0.068\,\mathrm{arc\,sec/(km\,s^{-1})}$. This position of the center of the nebula is puzzling. The vector from our center of the nebula to the pulsar points almost exactly opposite to the proper motion vector multiplied by time since explosion found by Caraveo and Mignani~\cite{c5}. This could be explained if the nebula was moving with respect to us with twice the velocity of the pulsar, which seems unlikely since Crab is considered a galactic disk object. Therefore, we checked the assumption that the center of the nebula is where Caraveo Mignani's proper motion result would put it, assuming that the proper motion of the nebula is negligible. We constructed a movie rotating about this center. In this case wobbling of the 3D image is beyond any doubt. Note, however, that due to slit length limitations the radial distribution is cut at $\approx 120\,\mathrm{arc\,seconds}$ from the center,  where the intensity is not completely negligible. This should be taken as a warning that the obtained fit values may not yet have a very simple physical meaning. If we assume in (\ref{eq:trimble}) that T is the time elapsed since the supernova explosion, we obtain $3.1\,$kpc as the best value for the distance, somewhat larger than the accepted value. Finally, in Fig.~\ref{RadDist} we show the radial distribution of intensity (proportional to the power radiated in a line in the shell between $\rho$ and $\rho+\Delta \rho$) in the three line sets as a function of the "radial distance" $\rho=\sqrt{(\xi - \xi_c)^2+ (\eta-\eta_c)^2 + \alpha^2 (v-v_c)^2}$. The sharp rise at $\approx 70\, \mathrm{arc\,seconds}$ is followed by gradual and regular fall off.

Our 3D image suggests that the volume within $\approx 55\, \mathrm{arc\,seconds}$ of the center of the nebula is almost void of line emitters. A noteworthy exception is a very weak trail of spots on the line connecting the upper left and lower right brightest spot and going through the center of the nebula in the spectra of slit orientation $144^\circ$, which is the orientation of the X-ray and radio jet in the image (see Fig.~\ref{Slits}). These spots are less than 5\% of maximum intensity and are likely along the jet. A saturated image of this reduced spectrum is shown in Fig.~\ref{144Sat}. 

\placefigure{RadDist}

\section{CONCLUSION}

We obtained multiple long slit spectra of the Crab nebula in the light of the optical emission lines of $\mathrm{H}_\alpha$, [NII]$\lambda$6548,6583, and [SII]$\lambda$6716,6730. The simple assumption of ballistic expansion, expressed by (\ref{eq:trimble}) and suggested by the work of Trimble~\cite{c1} allows a 3D representation of the nebula, although through an arbitrary scale in the line of sight. Such a 3D representation suggests that line emitting regions are organized in a relatively thin shell, regardless of the line of sight scale. Therefore, we make the assumption that the shell is likely to be quasi spherical and adjust the unknown scale parameter accordingly. The resulting 3D image indeed appears to fill a spherical shell with an inner radius of $\approx 55\,\mathrm{arc\,seconds}$, as already shown by results of Lawrence et al. \cite{c11}. The shell is pierced by the jet in the SE direction and at about $75^\circ$ with respect to the line of sight. The center of this shell is well aligned with the two-dimensional projection of the center of the optical and X-ray continuum emissions. It is clearly displaced from the pulsar  but appears to be connected to it through the north-western extension of the jet. The 3D distribution of the optical continuum and the X-ray emission can not be determined, but the observed 2D projections of these are very similar and consistent with the idea that the optical and X-ray continuum come from the same, rather flat, oblate structure centered in the spherical shell. The optical continuum emission drops regularly with distance from the center of the nebula (Fig.~\ref{Continuum0036}), becoming weak when reaching the spherical shell.  Based on the absence of the [NII]5755 line, we estimate that the effective gas temperature in the shell is below 7500K. The filling of the $55\,\mathrm{arc\,seconds}$ shell seems patchy. We note that some bright spots are only arc seconds wide (cf. Fig.~\ref{Oba036x}) and possibly less than $16\,\mathrm{km\,s^{-1}}$ ``deep'' (Fig.~\ref{OzkiInSiroki}), so that we expect  that a more detailed high resolution 3D view would show that the shell is woven with delicate threads. Note that the small size of line emitting regions indicates high emission coefficients in the bright regions; in particular, we estimate that the region shown in Figs.~\ref{Oba036x} and \ref{OzkiInSiroki} is only about 0.02 ly thick which leads to the H$_\alpha$ emission coefficient of the order $j_{\mathrm{H}_\alpha}\rho ~\approx ~10^{-21}\mathrm{\,erg\,cm^{-3}\,sec^{-1}}$, which is an order of magnitude higher than the average emission coefficient at these wavelengths of the whole nebula including the synchrotron cloud. Hints of threads are suggested in the superimposition of the radio image by Velusamy et al. \cite{c3} on the optical image and are clearly seen in the newer radio image by Bietenholz et al.~\cite{c7} shown in Fig.~\ref{FourImages}. Note that optical filaments closely map the smoother radio loops. The jet, which is so prominent in X-rays and in radio, is marked also by optical continuum emission and seems to be threaded also by weak line emission (see Fig.~\ref{Grobi036}, in particular at the $144^\circ$ orientation and Fig.~\ref{144Sat}). Line emission from regions where the jet hits the $55\,\mathrm{arc\,seconds}$ shell is particularly strong - the upper left brightest spot in the reduced spectrum $\psi_s=144^\circ$ and also $\psi_s=162^\circ$ of Fig.~\ref{Grobi036}.
 
In concluding, we would like to call attention to two observations: The roundness and average regularity of the spherical shell as expressed by the sharpness and simplicity of the radial distribution of line emitting regions in Fig.~\ref{RadDist} is quite remarkable. The delicate radiating tubes or blobs filling this relatively thin shell closely follow radio loops on large scale,  even if, as emphasized by Sankrit et al.~\cite{c34} and Hester et al.~\cite{c33}, on short scale they are dominated by the magnetic Rayleigh-Taylor instabilities. The 3D image, even if it still lacks in detail, suggests that the filaments are wound on the shell along long smooth threads, possibly following the magnetic field lines.

\acknowledgments
This work was done as a collaboration of the Instituto Nacional de Astrof\'{\i}sica, \'Optica y Electr\'onica (INAOE) and the University  of Ljubljana and was supported in part by CONACyT grant 25539E and by the Ministry of  Science, Education and Sport of the Republic of Slovenia under grant number 1554-501. The essential support to this project by the technical staff of the OAGH in Cananea is appreciated. The movie production is a courtesy of ARTREBEL9.

\newpage
\begin{figure}
\caption{\label{Oba036x}(a) High and (b) low resolution of the reduced spectrum for the slit angle $36^\circ$. Three 30 minute exposures for the high resolution spectrum were obtained on 2002 January 8 and five 20 minute for the low resolution spectra were obtained on 2002 January 6. High resolution spectrum covers the region between $6406\mathrm{\AA}$ and $6842\mathrm{\AA}$ ($\mathrm{0.43\AA /pixel}$) and the low resolution spectrum covers wavelengths between $5387\mathrm{\AA}$ and $7054\mathrm{\AA}$ ($\mathrm{1.6\AA /pixel}$). The vertical scale is 560 pixels $\times$ 0.46 arc seconds per pixel in both cases, but only pixels between $10<m<537$ are properly exposed in the high resolution spectrum and between $43<m<545$ for the low resolution spectrum. Continuum is calibrated in the brightened vertical region centered at 6000$\mathrm{\AA}$ shown in the lower half of the figure (b).}
\end{figure}

\begin{figure}
\caption{\label{Super036}Low (green) and high (red) resolution spectra in the spectral region between $6415$ and $6814~\mathrm{\AA}$ for the slit angle $36^\circ$ in the vicinity of the [NII]$\lambda$6548,6583 doublet, H$_\alpha \lambda$6563, and [SII]$\lambda$6716,6730 doublet suitably rescaled and superimposed.}
\end{figure}

\begin{figure}
\caption{\label{Identi}Trace $S_r(\lambda, m)$ of the spectrum $m = 440\,\textrm{pixels}$ (along the line in Fig.~\ref{Oba036x}b). Some important lines are marked; the rest positions with respect to local lamp calibration are shown by gray vertical lines and parallel  blue and red lines indicate blue/red shifted positions by $-1047\,\mathrm{km\,s^{-1}}$ and $1729\,\mathrm{km\,s^{-1}}$ respectively. Absolute velocity calibrations in low resolution spectra are not better than $\approx100\,\mathrm{km\,s^{-1}}$, but comparisons like the one in Fig.~\ref{Super036} show that calibration shifts are the same, to within the resolution of the high resolution spectra, in the whole region of [NII] and [SII] lines of interest. The scale on the ordinate axis is in photon counts per pixel per 20 minutes.}
\end{figure}

\begin{figure}
\caption{\label{Continuum0036}Continuum intensity at 6000\AA\ along the $36^\circ$ slit (see Fig.~\ref{Oba036x}b). The abscissa ticks are at 10 arc seconds spacing, and the ordinate shows the intensity as calibrated according to \cite{c8, c9}. $1\times 10^{-15}\mathrm{erg\,sec^{-1}\,cm^{-2}\,\AA^{-1}\,arcsec^{-2}}$ correspond to approximately 10 photon counts per pixel per 20 minutes. The dotted line indicates the continuum interpolation and the dot is placed at the intensity corresponding to the interpolated nebular continuum at the position of the pulsar, where calibration is applied.}
\end{figure}

\begin{figure}
\epsscale{0.5}
\caption{\label{Slits}Slit positions with respect to the nebula and their reconstruction. The slit orientation is as read from the dial on the spectroscope. The value of $m$ at the pulsar trace in the spectrum is taken as reference. The coordinate axis X ($\xi$) points to the West (right) and Y ($\eta$) points to the North (up). Star traces in the spectrum are noted and the corresponding distances to the reference trace are marked as points along a line with a given orientation. Such points (as for example the three black dots in the bottom image for slit orientation $0^\circ$) are parallelly transported until they correspond with some stellar feature in the image. The background image is a combination of three V, R, I filter images obtained at the Guillermo Haro observatory in Cananea by \v Cade\v z et al. \cite{c31} in the stroboscopic mode to enhance the brightness of the pulsar.}
\end{figure}

\begin{figure}
\caption{\label{OdstGrobi} Two successive subtractions of Gaussian multiplets at remaining maximums of the velocity weight function. In the three parts of the figure exemplary sections of low resolution spectra around the H$_\alpha$, [NII] and [SII] lines are shown as border lines between gray and white. The first subtraction leaves the red spectral trace, and the black line marks what remains after the second subtraction. The two right hand side [NII], H$_\alpha$, and [SII] quintuplets are clearly separated. Noise in the spectrum remaining after the first subtraction is roughly consistent with photon counting noise. The remaining weak multiplet (red) is again subtracted out to a line whose noise is consistent with the photon counting noise. The middle spectrum shows an example where a strong and weak [NII] and H$_\alpha$ triplets almost overlap. The first subtraction is possibly marginally worse than in the previous case. The left spectrum shows a case where the lines [NII] and H$_\alpha$ lines belonging to different triplets are superimposed. The automatic procedure makes no attempt to disentangle them. In the first step everything that is at the $v_{max}$ position of the velocity weight function for [NII] and H$_\alpha$ is subtracted. The second subtraction finds only a strong [SII] doublet and an H$_\alpha$ line.}
\end{figure}

\begin{figure}
\begin{center}
\end{center}
\caption{ \label{PraviInSinteticni}The $\psi=90^\circ$ low resolution spectrum and its reconstruction from velocity components. For color coding, abscissa and ordinate range on the reconstruction see Fig.~\ref{Grobi036}.}
\end{figure}

\begin{figure}
\caption{\label{OzkiInSiroki}Trace $S_r(\lambda, m)$ of the spectrum along the line in Fig.~\ref{Oba036x}a. The [SII] doublet at the red-shift $-$115 km/s is  unresolved and so is the weak H$_\alpha$ line, but the corresponding [NII] doublet is wide; another [SII] doublet at $814\,\mathrm{km\,s^{-1}}$ (to the right) is clearly about $150\,\mathrm{km\,s^{-1}}$ wide. Note that our spectral resolution of $16\,\mathrm{km\,s^{-1}}$ is not sufficient to show thermal broadening of H$_\alpha$ or any other line at expected temperatures of 7500K.}
\end{figure}

\begin{figure}
\epsscale{0.35}
\caption{\label{Grobi036} Reduced low resolution spectra for all slit orientations ($\psi_s~=~0^\circ ,~18^\circ~\dots ~162^\circ$- numbered in red). The intensity $w_i(\rho,v)$,  coded according to the color code shown by triangles, is shown as a function of velocity (abscissa: $v$ with ticks at 100$\,\mathrm{km\,s^{-1}}$) and position along the slit (ordinate: $\tilde m$, major ticks spaced by 1 $\mathrm{arc~min}$). The positive axis of the velocity coordinate points away from the observer and the positive axis of the slit coordinate points toward the numbers in Fig.~\ref{Slits}. The large orange cross indicates the true origin of the coordinate system centered on the pulsar at $v=0$, cf. eq. \ref {Coord}. The light gray ring with a dot at its center indicates the projection of the spherical shell volume ($R_{in}=65~ \mathrm{arc~sec}$, $R_{out}=90~ \mathrm{arc~sec}$) found by the fit. The colors are assigned with respect to the "average bright spot mixture of intensities", which is coded gray (center of the triangle, $\lbrace \mathrm{H:N:S} \rbrace$=$\lbrace 0.18: 0.42: 0.40\rbrace$; in the average hydrogen is much stronger), while the edges of the triangles correspond to pure H, N or S as indicated on top of the figure. Intensity is coded relative to each reduced spectrum as fraction of the maximum intensity of the given reduced spectrum. For respective maximum values see text. The minimum intensity seen in the electronic version of the image is about 5\% of the maximum intensity, which is significantly above noise, remember, however, that due to line interference the deconvolution is not unique; compare Fig.~\ref{PraviInSinteticni}.}
\end{figure}

\begin{figure}
\epsscale{0.35}
\caption{\label{FiniPr036}Same as Fig.~\ref{Grobi036} for the high resolution spectra. The S/N for this spectroscopy is quite low (see text), so that some specks are due to noise fluctuations, in particular the trace $\psi=18^\circ$ is the noisiest since it has the lowest maximum intensity (see text). This spectroscopy is mainly useful in conjunction with low resolution spectroscopy for better velocity calibration. Compare also with Fig.~\ref{Super036}.}
\end{figure}

\begin{figure}
\epsscale{0.35}
\caption{\label{144Sat} A saturated reduced low resolution spectrum for the slit orientation $\psi_s=144^\circ$ showing the trail of the jet. The color triangle corresponds to the intensity 5\%. The signal along the jet is up to 25\% of maximum intensity, yet it is important to realize that the spectrum in this region is crowded, so that some line confusion is probable. Higher resolution and better S/N spectroscopy of  this region is required.}
\end{figure}

\begin{figure}
\epsscale{0.50}
\caption{\label{FourImages}Crab nebula image reconstruction from the spectral data. Top left: a blend of optical and X-ray image, Weisskopf at al. \cite{c30};top center:radio image, Bietenholz et al. \cite{c7}; top right:  optical image with the dynamical center of the nebula and the projection of the main line emitting  lobe; bottom left: 6000\AA  \ continuum; bottom right: [NII] (green), H$_\alpha$ (red), [SII] (blue) spectral components.}
\end{figure}

\begin{figure}
\epsscale{0.50}
\caption{\label{Movie}An image (a single movie frame) of the Crab nebula from a different perspective. The observer on the Earth is in the direction of the $-Z$ axis. The color coding is explained in Sec.~\ref{image}.}
\end{figure}

\begin{figure}
\begin{center}
\end{center}
\caption{ \label{RadDist}Radial distribution of intensity in the three line sets as a function of the ``radial distance'' $\rho$.}
\end{figure}

\end{document}